\documentclass[twocolumn,showpacs,aps,floatfix,prl]{revtex4}
\usepackage{graphicx}
\def\la{\langle}
\def\ra{\rangle}
\def\beq{\begin{equation}}
\def\eeq{\end{equation}}
\def\beqa{\begin{eqnarray}}
\def\eeqa{\end{eqnarray}}

\begin{document}
\title{The role of initial state reconstruction in short and long time deviations\\ from exponential decay.}
\author{J. G. Muga}
\email{jg.muga@ehu.es}
\affiliation{Departamento de Qu\'\i mica-F\'\i sica, Universidad del Pa\'\i s Vasco, Apdo. 644, 48080
Bilbao, Spain}
\author{F. Delgado}
\email{qfbdeacf@lg.ehu.es}
\affiliation{Departamento de Qu\'\i mica-F\'\i sica, Universidad del Pa\'\i s Vasco, Apdo. 644,
48080 Bilbao, Spain}
\author{A. del Campo}
\email{qfbdeeca@lg.ehu.es}
\affiliation{Departamento de Qu\'\i mica-F\'\i sica, Universidad del Pa\'\i s Vasco, Apdo. 644,
48080 Bilbao, Spain}
\author{G. Garc\'\i a-Calder\'on}
\altaffiliation{Permanent address: Instituto de F\'{\i}sica,
Universidad Nacional Aut\'onoma de M\'exico,
Apartado Postal {20 364}, 01000 M\'exico, D.F., M\'exico}
\email{gaston@fisica.unam.mx}
\affiliation{Departamento de Qu\'\i mica-F\'\i sica, Universidad del Pa\'\i s Vasco, Apdo. 644,
48080 Bilbao, Spain}

\date{\today}

\begin{abstract}
We consider the role of the reconstruction of the initial state in the deviation from exponential decay at short and long times. The long time
decay  can be attributed to a wave that was, in a classical-like, probabilistic sense, fully outside the initial state or the inner region at
intermediate times, i.e., to a completely reconstructed state, whereas the decay during the exponential regime is due instead to a
non-reconstructed wave. At short times quantum interference between regenerated and non-regenerated paths is responsible for the deviation from
the exponential decay. We may thus conclude that state reconstruction is a ``consistent history'' for long time deviations but not for short
ones.
\end{abstract}
 \maketitle

%%%%%%%%%%%%%%%%%%%%%%%%%%%%%%%%%%%%%%%%%%%%%%%%%%%%%%%%%%%%%%%
%\section{Introduction}

{\it Introduction}. The decay from unstable quantum states is a recurrent and important topic in several fields of physics. The exponential
decay, which is by far the most common type, is surrounded by deviations at short and long times \cite{Khalfin57}.
%, and Gamow developed a theory that accounted for it.
%The existence of deviations was pointed out, first and mostly, theoretically, at both early %and long times with respect to the dominant,
%intermediate exponential regime \cite{Khalfin57}.
The short-time deviations have been much discussed, in particular in connection with the Zeno effect, and experimental observations have been
performed quite recently \cite{st}.
%We shall however concentrate here on the standard and much
%more common case of an exponential decay at intermediate times, surrounded by short-time %and long-time deviations.
An exactly solvable analytical model for a single resonance that describes these regimes follows for a decaying particle of mass $m$ from some
general considerations of the survival amplitude \cite{gcrr01},
\beqa A(T) = \la \psi_0|e^{-iHT/\hbar}|\psi_0\ra =\frac{i}{2\pi}\int_C \, e^{-ik^2T/\hbar} A(k)\,dk, \eeqa
(The units employed are $\hbar=2m=1$.), where the contour $C$ goes from $c+i\infty$ to $ic+\infty$, with $c >0$ a small constant,
 along the upper half complex $k$-plane
and $\psi_0$ is the initial state. If $A(k)$ is given by a single resonance, $A(k)= 2k [ D_r/(k-k_r)+D_r^*/(k+k_r^*)]$, and the input to
evaluate Eq. (\ref{at}) is the value of the complex pole $k_r^2=\varepsilon_r-i\Gamma_r/2$. It follows from the conditions $ A(T=0)=1$ and
$A(k=0)=0$ that \cite{gcrr01},
\beq\label{at} A(T)=\frac{1}{2}D_r e^{-y_r^2}-\frac{1}{2} [D_r\,w(-y_r)+D_r^*\,w(y_{-r})] \eeq
where $D_r=k_r/{\rm Re}\,(k_r)$, $ y_r=-{\rm exp}(-i\pi/4)k_rT^{1/2}$, $ y_{-r}={\rm exp}(-i\pi/4)k_r^*T^{1/2}$ and the function $w(z)=
\exp(-z^2) {\rm{erfc}}(-iz)$. The second term in Eq. (\ref{at}) dominates at short and long times compared with the lifetime. For other
analytical or semianalytical examples see \cite{gcmm95,MWS-AP96,MWS-EPL96}, which show that in general the exponential decay may be  related to a pole
residue and the deviation to a line integral resulting from a contour deformation along a steepest descent path in the complex momentum plane.
The long time deviations may also be associated with the boundedness of the spectrum \cite{Khalfin57} and the short time deviations with the
existence of the energy moments \cite{MWS-EPL96}, but clearly a more physical and intuitive understanding is desirable \cite{FG72,DN04}. In this
respect Fonda and Ghirardi in 1972 \cite{FG72} put forward an interesting connection following Ersak's previous insight \cite{Ersak69}: ``{\it
Concluding, we can say that the physical mechanism producing deviations from the exponential decay law of an unstable system is therefore
provided by the regeneration of the initial unstable state. In this regeneration process, a very decisive role is played by the decayed states,
which, as a consequence of the evolution of the system, are able to reconstruct partially the initial unstable state through a process of
rescattering}'' \cite{FG72}. While that paper was centered about long time deviations, in fact the arguments given are also valid for short
times, or arbitrary deviations \cite{FGR78}, as we shall see. Let us decompose the evolution up to the final time $T$ into different ``paths''
with the aid of the following complementary projectors
\beq P=|\psi_0\ra\la\psi_0|,\; Q= 1-P, \eeq
at an intermediate time $t$:
\beqa A(T)&=&\la \psi_0|e^{-iH(T-t)/\hbar}(P+Q)e^{-iHt/\hbar}|\psi_0\ra \nonumber
\\
&=&A(T-t)A(t) \nonumber
\\
\label{first} &+&\la \psi_0|e^{-iH(T-t)/\hbar}Qe^{-iHt/\hbar}|\psi_0\ra. \eeqa
Except for the last term, $A(T)$ would be exponential \cite{Ersak69,FG72}. Since the last term involves a path which goes out of the initial
state at the intermediate time, the deviation from exponential decay is attributed to state reconstruction.

The generality of this argument, however, diminishes somehow its explanatory power, and we may rightly wonder what is the difference, if any,
between the physical origin of short and long time deviations, in particular because some of the mathematical arguments provided to explain the
deviations are different for long (Paley and Wiener theorem) and short times (behaviour at the origin, and existence of energy moments). It also
seems reasonable to demand a quantitative rather than qualitative description of the reconstruction process. In addition, a key conceptual
question should be addressed: Since Eq. (\ref{first}) is an equation for {\it amplitudes}, the language used for interpreting it is possibly not
fully appropriate, and has to be made more precise. As it is well known, Feynman paths interfere, so that, in general, regarding a sequence of
events (a history) of the system previous to a final or measured event as something that has actually occurred is not justified, in the sense
that probabilities cannot be assigned consistently, in general, to the set of alternative histories. This has been emphasized lately by the
consistent histories interpretation, in which a time sequence of events can only be considered a ``consistent'' (classical-like) history under
specific conditions, in a nutshell when the alternative paths do not interfere \cite{CH}.
%Rather than dealing with complex and interfering amplitudes,
%the classical-like world of events
%is described by the probabilities of alternative histories.

By applying the resolution of the identity $1=P+Q$ at an intermediate time $t$, the survival probability is decomposed, using Eq.(\ref{first})
and its complex conjugate, as
\beq \label{deco} S(T)=|A(T)|^2=[PP]_t+[PQ]_t+[QP]_t+[QQ]_t, \eeq
where we have introduced the notation,
\beqa [XY]_t&=&\la \psi_0|e^{-iH(T-t)/\hbar}Xe^{-iHt/\hbar}|\psi_0\ra \nonumber
\\
&\times&\la \psi_0|e^{iHt/\hbar}Ye^{iH(T-t)/\hbar}|\psi_0\ra, \eeqa
(or $[XY]$ for short) in which $X$ and $Y$ can be any of the projectors $P$ or $Q$. It is a simple exercise to see that all the terms that
contribute to the survival probability in Eq. (\ref{deco}) can be reduced to products of survival amplitudes for different times,
\beqa [PP]_t&=&|A(T-t)A(t)|^2,
\nonumber\\
{[}PQ{]}_{t}&=&A(T-t)A(t)A^*(T)-|A(T-t)A(t)|^2,
\nonumber\\
{[}QP{]}_t&=&A(T-t)^*A(t)^*A(T)-|A(T-t)A(t)|^2,
\nonumber\\
{[}QQ{]}_t&=&|A(T)|^2-A(T-t)A(t)A^*(T),
\nonumber\\
&-&A(T-t)^*A(t)^*A(T)+|A(T-t)A(t)|^2. \eeqa
The diagonal terms $[PP]_t$ and $[QQ]_t$ have simple ``event'' interpretations: for $[PP]_t$ the particle is initially in $|\psi_0\ra$, it is
again in $|\psi_0\ra$ at the intermediate time $t$, and ends up in the same state at $T$; for $[QQ]_t$ the particle begins and ends in
$|\psi_0\ra$ but it is in the complementary subspace spanned by $Q$
at the intermediate time $t$. This is therefore the term associated with initial state
reconstruction. The nondiagonal terms, however, are quantum interferences without classical counterpart. Thus the ``histories'' represented by
$[PP]_t$ and $[QQ]_t$ terms are only ``consistent'' if the contribution from the interference terms is negligible, i.e., if the survival
probability may be attributed essentially to these diagonal terms only. In other words, when the histories are consistent we may plainly say
that the events have indeed happened in one or the other way with certain probabilities, without the need to invoke virtual paths and complex
amplitudes. There remains of course an arbitrariness in the definition of ``negligible'', since the interference terms are often small compared
to the others but rarely zero. One should accept, clearly, that the ``consistency'' or ``classicality'' of the histories is not absolute and
sharply defined  but a gradual quality that may however be precisely quantified.

%``State reconstruction´´, as a classical-like history, is associated with the
%the term $[QQ]_t$. We have performed calculations of this term for
%different models.

{\it The models}.
%
%\subsection{One resonance model}
%
The exact single resonance model for decay as the one discussed before, see \cite{gcrr01}, is possibly the simplest one for decomposing the
survival since $A(t)$ reduces to a known function. One pays a price though, as the single resonance model does not contain an explicit form of the
Hamiltonian or the initial state. The numerical calculations presented here correspond to a more explicit delta-function Hamiltonian model
described below. Note, however, that all basic properties of the survival decomposition found for the delta model have also been reproduced for
the simple single resonance model, although the results for the later are not shown to avoid unnecessary repetitions.
%
%\subsection{Delta function model}
%

In the delta function model the particle is confined to the half-axis $x>0$ with boundary condition $\psi(x=0)=0$, and subjected to a
delta-function potential,
\beq \label{v} V=\eta\delta(x-1). \eeq
%
%To solve this problem we consider a dimensionless Schr$\ddot{\rm{o}}$dinger equation
%
%\beq
%$ i\partial\psi/\partial t=-\partial^2\psi/\partial x^2  + V(x) \psi $
%\eeq
%that may be converted into a dimensional one making use of a characteristic length $L$ for the delta position, energy $\hbar^2/(2mL^2)$ and time
%$2mL^2/\hbar$.
%
In all calculations the initial state is the ground state of the infinite well,
%
%\beq
$ \psi(x,t=0)=\sqrt{2}\sin(\pi x). $
%\eeq
%
The evolved wavefunction $\psi(x,t)$ is calculated by using the resolution of the identity based on the eigenfunctions of the stationary
Schr$\ddot{\rm o}$dinger equation, delta-normalized in $k$ space. Since their overlap with the initial function can be calculated in the form of
an explicit function, $\psi(x,t)$ is reduced to a single integral which is performed numerically. The calculation may be also carried out by
expanding $\psi(x,t)$ in terms of the resonant-state basis \cite{gcmm95}.

%\subsection{Delta function model with external absorbing potential}
%
Consider now a model that represents the effect of a detector by adding to the potential
in Eq. (\ref{v}), an external absorbing potential $-iV_0$ for $x\ge 1$ \cite{cplas}.
%
%\beq
%V'=\Bigg\{\begin{array}{ll}
%0&0\le x\le 1
%\\
%-iV_0& x>1,\hspace{.5cm} V_0\ge 0
%\end{array}
%\eeq
%
The solutions of the stationary Schr$\ddot{\rm o}$dinger equation which are right eigenvectors of the Hamiltonian with eigenvalue $E=q^2-iV_0$
can be written as
\beq \phi_q(x)=\frac{1}{(2\pi)^{1/2}}\Bigg\{
\begin{array}{ll}
C_1(e^{ikx}-e^{-ikx}),&0\le x\le 1
\\
e^{-iqx}-{\cal S}e^{iqx},& x\ge 1
\end{array}
\label{form} \eeq
where
%
%\beq
$ k=(q^2-iV_0)^{1/2} $
%\eeq
%
is the wavenumber inside, and $q$ the wavenumber outside. For scattering-like solutions, $q$ is positive. Note the two branch points of $k$ in
the complex $q$ plane. We shall take the branch cut joining these points. Similarly, the root in $q=(k^2+iV_0)^{1/2}$ is defined with a branch
cut joining the two branch points in the $k$ plane.
%The matching conditions are
%
%\beqa
%\phi_q(1^-)&=&\phi_q(1^+),
%\\
%\phi_q'(1^+)-\phi_q'(1^-)&=&\eta\phi_q(1),
%\eeqa
%
%where the prime denotes spatial derivative, and the superscript
%$\pm$ the side from which the limit is taken, $+$ from the right
%and $-$ from the left.
The matching conditions determine the amplitudes $C_1$ and ${\cal S}$. The right eigenvectors of a complex Hamiltonian have biorthogonal
partners $\widehat{\phi}_q(x)$, which are left eigenvectors of $H$ with the same eigenenergy $E$ \cite{cprev}. Alternatively they may also be
regarded as right eigenvectors of $H^\dagger$ with eigenvalue $E^*$.
%They take the form
%\beq
%\widehat{\phi}_{q'}(x)=\frac{1}{(2\pi)^{1/2}}\Bigg\{
%\begin{array}{ll}
%\widehat{C}_1(e^{ik'x}-e^{-ik'x}),&0\le x\le 1
%\\
%e^{-iq'x}-\widehat{S}e^{iq'x},& x\ge 1
%\end{array}
%\label{form}
%\eeq
%
%with $q'=q$ and $k'=k^*$.

Aside from the scattering solutions, there may be localized solutions for complex $q_j$ and positive imaginary part,
% The right eigenvectors
%take the form
%
%\beq
%u_j(x)=\frac{1}{(2\pi)^{1/2}}\Bigg\{
%\begin{array}{ll}
%C_j(e^{ik_jx}-e^{-ik_jx}),&0\le x\le 1
%\\
%D_je^{iq_jx},& x\ge 1
%\end{array}
%\eeq
%
%where the discrete wavenumbers are found by solving
%
%\beq
%k_j+(-iq_j+2m\eta/\hbar^2)\tan(k_j)=0
%\eeq
%
%for Im$(q_j)>0$. (This is the same equation for the poles
%of $S$.) The subscript $j$ is used to label them.
normalized with respect to their biorthogonal partners
%(having generally complex wavenumbers $k_j'=-k_j^*$ and $q_j'=-q_j$)
as
%
%\beq
$ \la \widehat{u}_j|u_i\ra=\delta_{ij}. $
%\eeq
%
%where $\delta{ij}$ is a Kronecker delta.
The time evolution in this case is obtained from the biorthogonal resolution of the identity,
%
%\beq
$ 1=\sum_j |u_{j}\ra\la\widehat{u}_j| +\int_0^\infty dq |\phi_q\ra\la \widehat{\phi}_j|. $
%\eeq
%
%which we have checked by decomposing arbitrary states.
%
%
%
%
%
% ---------------- FIG. 1 BEGINS ----------------
\begin{figure}[t]
\begin{center}
\includegraphics[angle=-90,width=0.65\linewidth]{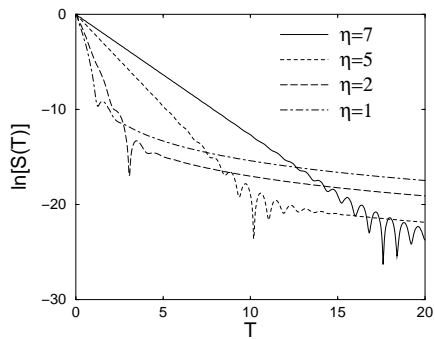}
\end{center}
\caption{\label{1}$\ln$ of the survival probability $S(t)$ for different delta function strength factors.}
\end{figure}
% ---------------- END FIG. 1 ----------------
%
%\section{Results and discussion}
%

{\it Results and discussion}. Typical survival curves for different $\eta$ factors are shown in Fig. \ref{1}. ($V_0=0$ for the time being.) Note
the clear transition between exponential and long time deviations (The short time deviation is not seen in the scale of the figure).
%
%\subsection{Long times}
%

\begin{figure}[t]
\begin{center}
\includegraphics[angle=-90,width=0.65\linewidth]{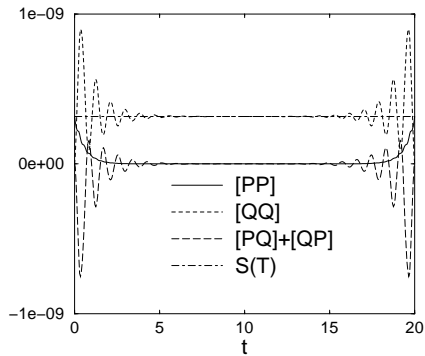}
\end{center}
\caption{\label{ltime} Survival
%Dominance of reconstructed term at long times:
probability $S(T)$ and its decomposition
% into non-reconstructed, $[PP]_t$, reconstructed, $[QQ]_t$,
%and interference contributions, $[PQ]_t+[QP]_t$,
as a function of the intermediate time $t$ for a final time $T=10$ in the long-time, non-exponential decay regime. $\eta=5$.}
\end{figure}

First we decompose the survival at a large time $T$ in which the decay is not exponential for different intermediate times $t$ according to Eq.
(\ref{deco}). Figure \ref{ltime} shows the clear dominance of the reconstructed term. The other terms are negligible near the middle
intermediate time point, $2.5 < t < 7.5$, so that the reconstruction is a consistent history, in fact the only significant one, in this range of
intermediate times, whereas the interference and non-reconstructed terms are not negligible for extreme intermediate times.
%
% ---------------- FIG. 2 BEGINS ----------------
% ---------------- END FIG. 2 ----------------
%

%\subsection{Long times: Nonescape probability}
%
No fundamental difference is observed if instead of the survival probability $S(T)$, the nonescape probability
$ N=\int_0^\infty dx\,|\psi(x,t)|^2 $
is decomposed
%
%
% ---------------- FIG. .. BEGINS ----------------
%\begin{figure}[t]
%\begin{center}
%\includegraphics[angle=-90,width=0.7\linewidth]{pdelta10p}
%\end{center}
%\caption{Dominance of reconstructed term at long times: Decomposition of the non-escape %probability into non-regeneration, $[PP]_t$,
%regeneration, $[QQ]_t$, and
%interference terms for a final time $T=10$ in the long-time,
%non-exponential regime. $\eta=2$.}
%\end{figure}
% ---------------- END FIG. .. ----------------
%
%
into terms $[PP]_t$, $[QQ]_t$ and interference contributions as before, but using now the new projector
%
%\beq
$ P=\int_0^1 dx\,|x\ra\la x|=\sum_{j=1}^\infty |E_j\ra\la E_j|, $
%\eeq
%
where $|E_j\ra$ are the eigenstates of the infinite well, and $Q=1-P$. In practice the sum is only taken up to a finite number of terms until
convergence is achieved. The calculation reduces then to evaluate matrix elements of the general form $\la E_n|e^{-iHt/\hbar}|E_{l}\ra$. The
resulting figure for $\eta=5$ and $T=3$, not shown, is almost identical to Fig. \ref{ltime}.

\begin{figure}[t]
\begin{center}
\includegraphics[angle=-90,width=0.65\linewidth]{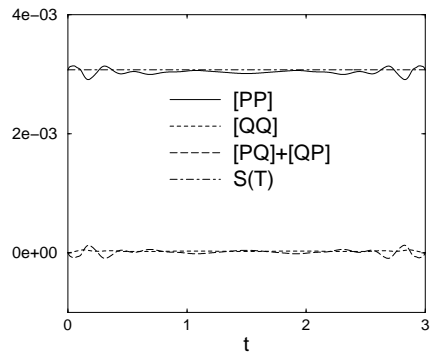}
\end{center}
\caption{\label{expo} Decomposition
%Dominance of the non-reconstructed term at exponential decay times:
of the survival probability
% into
%non-reconstructed, $[PP]_t$, reconstructed, $[QQ]_t$,
%and interference terms versus the intermediate time $0\le t\le T$
for a final time $T=3$ where exponential decay holds. $\eta=5$.}
\end{figure}

%\subsection{Exponential decay}
%
At variance with the long time case, in the region of exponential decay the   non-reconstructed term $[PP]_t$ dominates, see Fig. \ref{expo},
the other ones being negligible at all intermediate times.
%
% ---------------- FIG. 3 BEGINS ----------------
% ---------------- END FIG. 3 ----------------
%\subsection{short times}
%

\begin{figure}[t]
\begin{center}
\includegraphics[angle=-90,width=0.65\linewidth]{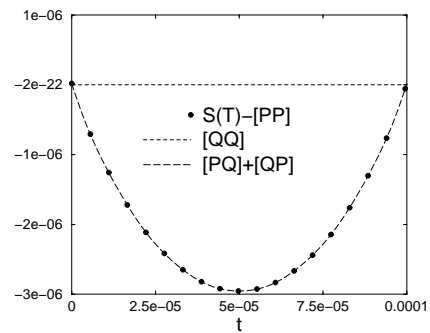}
\end{center}
\caption{\label{stime} Importance of the interference at short times: Decomposition of the survival probability for T=0.0001. $\eta=5$.}
\end{figure}

In the region of short time deviations the $[PP]_t$, non-reconstructed term dominates too. Nevertheless, there is a deviation from exponential
decay that must be explained by the  difference with the survival $S(T)-[PP]_t$. This difference essentially coincides, and it is numerically
undistinguishable in our calculations, with the interference term $[PQ]_t+[QP]_t$ for all intermediate times $t$, see Fig. \ref{stime}.
Therefore, initial state reconstruction in this regime cannot be regarded as a consistent history. It is an interference effect!
%
% ---------------- FIG. 4 BEGINS ----------------
% ---------------- END FIG. 4 ----------------
%
%\subsection{Effect of a detector}
%
% ---------------- FIG. 5 BEGINS ----------------
\begin{figure}[t]
\begin{center}
\includegraphics[angle=-90,width=0.65\linewidth]{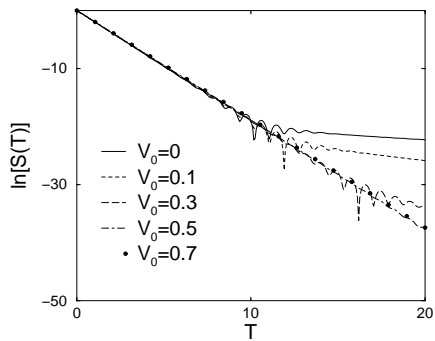}
\end{center}
\caption{\label{detector} Effect of a detector, represented by an imaginary potential $-iV_0$ outside the internal region limited by the delta.}
%The transition time between the exponential and long-time
%regimes increases with $V_0$, demonstrating the suppression of state reconstruction.}
\end{figure}
% ---------------- END FIG. 5 ----------------
%
% ---------------- FIG. 6 BEGINS ----------------
\begin{figure}[t]
\begin{center}
\includegraphics[angle=-90,width=0.65\linewidth]{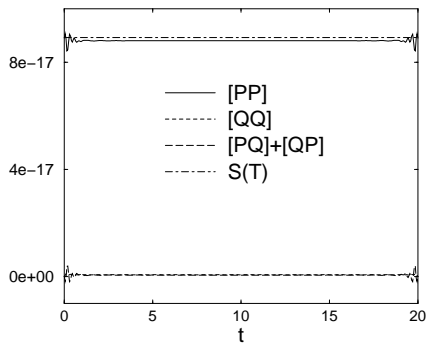}
\end{center}
\caption{\label{components} Dominance of the non-reconstructed $[PP]$-term in the survival when the detector is on. $\eta=5$, $V_0=1$. Compare
with Fig. \ref{ltime}.}
\end{figure}
% ---------------- END FIG. 6 ----------------

Fonda, Ghirardi and Rimini \cite{FGR78} pointed out that the state reductions associated with measurement interactions could hinder initial
state reconstruction and make the decay purely exponential \cite{FGR78}. They substantiated this idea with a measurement model producing state
reductions at random times. Several models have been used afterwards by different authors to show the effects of a detector in the exponential
decay and its deviations \cite{EG00,PL02}. Here we propose a complex imaginary potential as a simple way to represent atomic detection by
fluorescence induced by laser (this is justified in \cite{cplas}). Fig. \ref{detector} shows that the presence of a weak detector pushes the
transition between the exponential and long time deviation to larger times with increasing $V_0$, showing the gradual suppression of initial
state reconstruction.  This can also be seen explicitly in Fig. \ref{components}, where the dominant term in the decomposition is $[PP]$ instead
of $[QQ]$ as in Fig. \ref{ltime}. For the same values of $V_0$, however, the short time deviations are not affected because there is no enough
time for absorption to occur. Note that the slope of the long time deviation increases with $V_0$. In fact around $V_0\approx\Gamma$ it actually
disappears.(A more extensive analysis of this measurement model will be presented elsewhere, in particular for strong interactions, which
produce important changes in the life time, but are not so relevant for our present purpose.)
%
%\section{Conclusions}

There is in summary a significant difference between short and long time deviations from exponential decay regarding the role of state
reconstruction, which becomes a consistent history for long times but not for short times.

%%%%%%%%%%%%%%%%%%%%%%%%%%%%%%%%%%%%%%%%%%%%%%%%%%%%%%%%%%%%%%%%%%%%%%%%
%We thank ......... for comments.
\begin{acknowledgments}
This work has been supported by Ministerio de Educaci\'on y Ciencia (BFM2003-01003), and UPV-EHU (00039.310-15968/2004). G. G-C. acknowledges
also the financial support of DGAPA-UNAM (IN10803) and Ministerio de Educaci\'on y Ciencia (SAB2004-0010).
A.C. acknowledges financial support by the Basque Government (BFI04.479).\\
\end{acknowledgments}


\begin{thebibliography}{1}

\bibitem{Khalfin57} L. A. Khalfin, Zurn. Eksp. Teor. Fiz. {\bf 33}, 1371 (1957),
English translation: Sov. Phys. JETP 6 1053 (1958).

\bibitem{st} S. R. Wilkinson {\it et al.}, Nature (London) {\bf 387}, 575 (1997);
Q. Niu and M. Raizen, Phys. Rev. Lett. {\bf 80}, 3491 (1998).

%\bibitem{GW64} M. L. Goldberger and K. M. Watson, Phys. Rev. {\bf 136}, B1472 (1964).

%\bibitem{nonexp} T.Jittoh, S. Matsumoto, J. Sato, Y. Sato, and K. Takeda,
%Phys. Rev. A {\bf 71}, 012109 (2005).

\bibitem{gcrr01} G. Garc\'\i a-Calder\'on, V. Riquer, and R. Romo,
J. Phys. A {\bf 34}, 4155 (2001).

\bibitem{gcmm95} G.  Garc\'\i a-Calder\'on, J. L. Mateos and M. Moshinsky, Phys. Rev.Lett. {\bf 74}, 337 (1995);
G.  Garc\'\i a-Calder\'on, J. L. Mateos and M. Moshinsky, Phys. Rev.Lett. {\bf 90}, 028902-1 (2003).

%\bibitem{gc92} G. Garc\'{\i}a-Calder\'on in {\it Symmetries in Physics},
%edited by A. Frank and K. B. Wolf (Springer-Verlag, Berlin, 1992)p.252.

\bibitem{MWS-EPL96} J. G. Muga, G. W. Wei, and R. F. Snider,
Europhys. Lett. {\bf 35}, 247 (1996).

\bibitem{MWS-AP96} J. G. Muga, G. W. Wei, and R. F. Snider, Annals of Physics {\bf 252}, 336
(1996).

\bibitem{FG72} L. Fonda and G. C. Ghirardi, Il Nuovo Cimento {\bf 7A}, 180 (1972).

\bibitem{DN04} T. G. Douvropoulos and C. A. Nicolaides, Phys. Rev. A {\bf 69},
032105 (2004).

\bibitem{Ersak69} L. Ersak: Yad. Fiz. {\bf 9}, 458 (1969); English translation: Sov. J. Nucl. Phys. {\bf 9}, 263 (1969).

\bibitem{FGR78} L. Fonda, G. C. Ghirardi, and A. Rimini, Rep. Prog. Phys.
{\bf 41}, 587 (1978).

\bibitem{CH} R. B. Griffiths, J. Stat. Phys. {\bf 36}, 219 (1984);
R. Omn\`es, The Interpretation of Quantum Mechanics (Princeton University Press, Princeton, 1994); M. Gell-Mann and J. B. Hartle, Phys. Rev. D
{\bf 47}, 3345 (1993); J. J. Halliwell, in Time in Quantum Mechanics, ed. by J. G. Muga, R. Sala and I. L. Egusquiza (Springer, Berlin, 2002).

\bibitem{cplas} A. Ruschhaupt, J. A. Damborenea, B. Navarro, J. G. Muga
and  G. C. Hegerfeldt,
%``Exact and approximate
%complex potentials for
%modelling time observables''
Eur. Phys. Lett. {\bf 67}, 1 (2004).


\bibitem{cprev} J. G. Muga, J. P. Palao, B. Navarro, I. L. Egusquiza,
%``Complex Absorbing Potentials'',
Physics Reports {\bf 395}, 357 (2004).

\bibitem{EG00} B. Elattari and S. A. Gurvitz, Phys. Rev. {\bf 62}, 032102 (2000).

\bibitem{PL02} R. E. Parrot and J. Lawrence, Europhys. Lett. {\bf 57}, 632 (2002).
%persisence of exp. decay at long times

%103

%\bibitem{Panov} A. D. Panov, Phys. Lett. A {\bf 270}, 441 (1999).

%\bibitem{EM00} I. L. Egusquiza and J. G. Muga, Phys. Rev. A {\bf 62}, 032103 (2000).

\end{thebibliography}
\end{document}